# Absolute Energy Measurements with Superconducting Transition-Edge Sensors for Muonic X-ray Spectroscopy at 44 keV

**Daikang Yan[1,2] • Joel C. Weber[1,2] • Tejas Guruswamy[3] • Kelsey M. Morgan[1,2] • Galen C. O'Neil[1] • Abigail L. Wessels[1,2] • Douglas A. Bennett[1] • Christine G. Pappas[1,2] • John A. Mates[1,2] • Johnathon D. Gard[1,2] • Daniel T. Becker[1,2] • Joseph W. Fowler[1,2] • Daniel S. Swetz[1] • Daniel R. Schmidt[1] • Joel N. Ullom[1,2] • Takuma Okumura[1,4] • Tadaaki Isobe[4] • Toshiyuki Azuma[4] • Shinji Okada[5] • Shinya Yamada[6] • Tadashi Hashimoto[7] • Orlando Quaranta[3] • Antonino Miceli[3] • Lisa M. Gades[3] • Umeshkumar M. Patel[3] • Nancy Paul[8] • Guojie Bian[8,9] • Paul Indelicato[8]**

[1] National Institute of Standards and Technology, Boulder, CO 80305 USA
[2] University of Colorado Department of Physics, Boulder CO 80309 USA
[3] Advanced Photon Source, Argonne National Laboratory, Lemont, IL 60439, USA
[4] RIKEN, Wako, Saitama 351-0198, Japan
[5] Chubu University, Kasugai 487-8501, Japan
[6] Rikkyo University, Toshima City, Tokyo 171-8501, Japan
[7] Japan Atomic Energy Agency, Tokai 319-1195, Japan
[8] Laboratoire Kastler Brossel, Sorbonne Université, CNRS, ENS-PSL Research University, Collège de France, Case 74, 4, place Jussieu, F-75005 Paris, France
[9] Key Laboratory of Computational Physics, Institute of Applied Physics and Computational Mathematics, 100088 Beijing, China

**Abstract** Superconducting transition-edge sensor (TES) microcalorimeters have great utility in x-ray applications owing to their high energy resolution, good collecting efficiency and the feasibility of being multiplexed into large arrays. In this work, we develop hard x-ray TESs to measure the absolute energies of muonic-argon (μ-Ar) transition lines around 44 keV and 20 keV. TESs with sidecar absorbers of different heat capacities were fabricated and characterized for their energy resolution and calibration uncertainty. We achieved ~ 1 eV absolute energy measurement accuracy at 44 keV, and < 12 eV energy resolution at 17.5 keV.

Author 1 • Author 2 • Author 3

# 1 Introduction

The muon is an elementary particle with the same electric charge as the electron but 200 times more mass. When a muon is captured by a nucleus, a muonic atom is produced. Due to the large mass of the muon, the Bohr radius of the muonic atom is 1/200 as large and subsequently the muon experiences an electric field 40,000 times stronger than that of an electron in the normal atom. For this reason, the muonic atom is a good system to explore quantum electromagnetic dynamics (QED) under extremely strong electric fields. One way of studying the QED effect in the muonic atom is by measuring its characteristic x-rays and comparing their energies with theoretical predictions [1]. The measurement requires detectors that have both high collecting efficiency and high energy resolution [2], which makes the transition-edge sensor (TES) a good candidate.

In order to calculate QED caused energy shifts of several tens of eV, this project aims to measure the absolute energies of muonic argon (μ-Ar) deexcitation at 44 keV and 20 keV with ~ 1 eV level accuracy. In previous work [2], we designed "sidecar" TESs with different numbers of bars to adjust thermal sensitivity and absorbers of different sizes to adjust heat capacity. Saturation energies of 41 keV and 48 keV were obtained for TESs that have 4 pJ/K and 5 pJ/K absorbers, respectively. In the work described in this paper, we continue to use the same sidecar layout but increase the absorber heat capacity $C$ to further improve the dynamic range. Electroplated bismuth (Bi) is added to the absorber to boost the detector quantum efficiency (QE).

# 2 Experiment

The TESs are 90 μm × 225 μm rectangular Mo-Au bilayers with 44 nm thick Mo and 630 nm thick Au. They are fabricated with a hybrid additive-subtractive process [3]. The critical temperatures of the TESs are 95 mK, and the normal resistances are 6.3 mΩ. The absorbers are bilayers of 1.8 μm thick Au and 18 μm Bi. They are made into three sizes: 625 μm × 625 μm, 685 μm × 685 μm, and 740 μm × 740 μm. Their heat capacities attributed to the Au portion at 100 mK are calculated to be 5 pJ/K, 6 pJ/K, and 7 pJ/K, respectively. The TES size is very small compared to the absorber size, and the specific heat capacity of Bi has been reported to be two orders of magnitude smaller than that of Au [4]. Therefore, the total heat capacity of

# Title

these devices mainly comes from the Au layer in the absorber. For the convenience of naming, in this manuscript these prototypes are called 5 pJ/K, 6 pJ/K, and 7 pJ/K pixels, respectively. We test pixels with different heat capacities because the detector energy resolution value $\Delta E$ and the saturation energy $E_{sat}$ both increase with $C$ [2]. Given the fact that TESs are not perfectly linear even below $E_{sat}$, a larger $E_{sat}$ value usually means that the device has better linear performance at low energies. Because the absolute energy measurement accuracy improves with smaller $\Delta E$ and larger $E_{sat}$, it is crucial that proper $C$ is selected in order to balance $\Delta E$ and $E_{sat}$.

The microcalorimeters are placed in an adiabatic demagnetization refrigerator at 70 mK bath temperature and are biased at 10% normal resistance. Their signals are read out by microwave superconducting quantum interference device (SQUID) multiplexers via a pair of coaxial cables [5]. The SQUID bandwidth is 300 kHz, and the flux-ramp rate is 62.5 kHz. The characterization is carried out at Beamline-1BM-C, Advanced Photon Source, Argonne National Laboratory, USA. We use 78 keV synchrotron beam source to excite characteristic x-rays from Gd, Tb, and Dy foils. The three fluorescent materials are measured at the same time, so that their x-ray signals have the same drift and gain, and therefore can be corrected simultaneously. The emission energies of these fluorescent materials are shown in Table I [6].

Table I. Emission Energies of Fluorescent Target Materials. Numbers in parentheses are one standard deviation uncertainties referred to the last significant digits of the quoted values.

| Emission line | Gd [eV] | Tb [eV] | Dy [eV] |
|---|---|---|---|
| $K\alpha_2$ | 42309.30(43) | 43744.62(46) | 45208.27(49) |
| $K\alpha_1$ | 42996.72(44) | 44482.75(47) | 45998.94(51) |
| $K\beta_1$ | 48696.9(57) | 50382.9(61) | 52119.7(65) |

## 3 Result and Discussion

Figure 1 shows the linearity feature of the three types of pixels. Each color represents fluorescence data on a single pixel. Since the transition edge of the TES is not perfectly linear [7], the pulse shapes created by photons of different energies are slightly different. As the saturation energy of the TES is exceeded, the pulse peak flattens, and the peak amplitude does not increase with energy as much as the pulse area does, causing the scatter plot of the pulse peak amplitude vs. pulse area to bend horizontally [2]. As shown in Fig. 1, the 5 pJ/K pixel starts to saturate around 50 keV, while the 6 pJ/K and 7 pJ/K pixels show nonlinear yet unsaturated behavior within the 52 keV measured range.

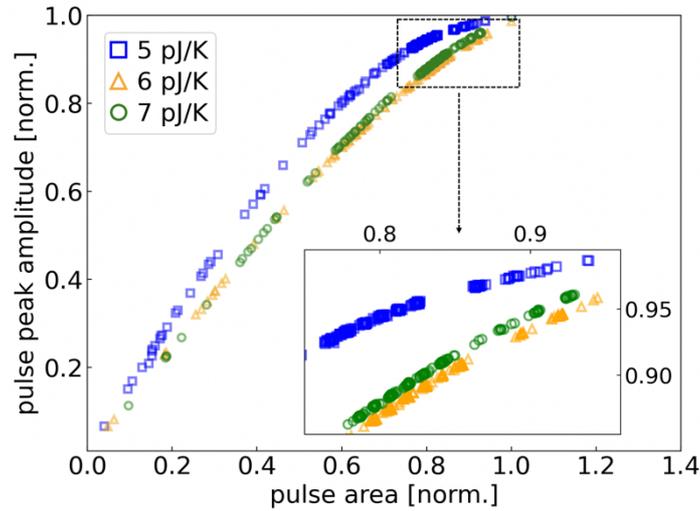

**Fig. 1** Pulse peak amplitude vs. pulse area for the 5 pJ/K (*blue square*), 6 pJ/K (*orange triangle*), and 7 pJ/K (*green circle*) single pixels in the energy range of 0 to 52 keV. The x and y axis values are normalized by their maxima, respectively. The inserted plot in the lower right corner enlarges the 42 keV to 52 keV emission lines of the fluorescence targets. Data points far below 42 keV are background fluorescence from materials in the absorber and sample stage. (Color figure online)

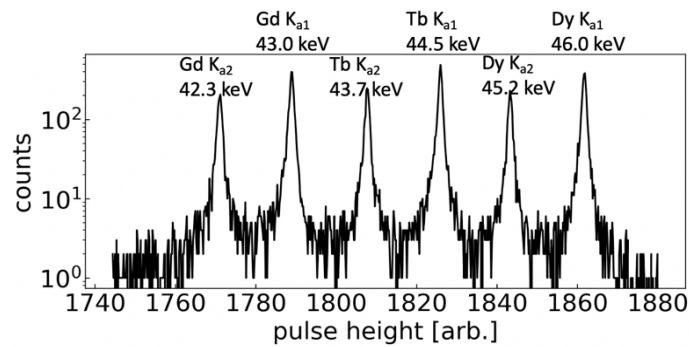

**Fig. 2** Optimally filtered pulse height histogram of the Gd, Tb, Dy $K\alpha_1$, $K\alpha_2$ emission lines on a 6 pJ/K pixel. The peak energies are annotated in the plot.

**Title**

The goal of energy calibration is to establish the relation between the photon energy $E$ and the pulse height $P$. We calculate the pulse heights with the optimal filter method [8], followed by the correction for signal gain variation that is caused by cryostat temperature drift [10]. Figure 2 shows the histograms of pulse heights of the six $K\alpha_1$, $K\alpha_2$ emission lines on a 6 pJ/K pixel. Owing to the thick Bi layer used in the absorber, a portion of photon events suffer partial energy loss [13, 14], which causes an exponential tail on the low-energy side of the measured spectrum. However, the low-energy tail effect is not observable in this measurement, due to the low statistics (~ 3000 photons per peak) and the large peak width of the histograms. The histograms can be well fitted with a Lorentzian function, and their center pulse heights are calculated.

There are several methods to construct the $E$ vs. $P$ relation. In this work, instead of interpolating the $E$ vs. $P$ data directly, we define the pulse gain as $g = P/E$ and fit with $P$ vs. $g$ values [11]. The 6 pJ/K and 7 pJ/K data are fitted with linear functions. The 5 pJ/K data shows a stronger nonlinear trend in the energy range of interest, and is therefore fitted with a quadratic function. In the calibration process, anchor points above and below the target energy are needed. Because the Tb $K\alpha_1$, $K\alpha_2$ lines are close to the 44 keV target project energy, we use Gd and Dy $K\alpha_1$, $K\alpha_2$ emission lines as anchor points and interpolate to find Tb $K\alpha_1$, $K\alpha_2$ line energies. The $K\beta_1$ lines are not used in the calibration because of the large uncertainty (~ 6 eV) in the reference values. Table II shows the deviation from the reference data of the calibrated energies of the three types of TESs. The energy deviation of the 6 pJ/K and 7 pJ/K pixels are within the ~ 1 eV calibration error. These two devices' absolute energy measurement accuracies are sufficient for the 44 keV experiment.

Table II. Tb $K\alpha_1$, $K\alpha_2$ line energy deviation on the three types of TESs.

| energy deviation [eV] | Tb $K\alpha_1$ 44.5 keV | Tb $K\alpha_2$ 43.7 keV |
|---|---|---|
| 5 pJ/K | 2.11(116) | 1.00(122) |
| 6 pJ/K | 0.38(76) | 0.13(84) |
| 7 pJ/K | 0.01(103) | 0.24(107) |

The $\Delta E$ of the TESs, defined as the full width at half maximum (FWHM), are extracted from the calibrated energy data. Fitting the calibrated Tb $K\alpha_1$ energy peak by a Voigt function (Fig. 3) with the 28.12 eV intrinsic Lorentzian line width [12], the energy resolutions at 44.5 keV of the 5 pJ/K, 6 pJ/K, 7 pJ/K pixels are calculated to be 20.42 ± 1.39 eV, 20.30 ± 1.27 eV, 17.69 ± 1.11 eV, respectively. Theoretically, $\Delta E$ of a TES is proportional to the square root of $C$. However, such positive correlation is not obtained in our measurement for two reasons: 1. the noise across different TES channels varies, and we happen to have smaller noise on the 7 pJ/K, which leads to its better energy resolution; 2. comparing the 5 pJ/K and 6 pJ/K pixels, which have similar noise levels, the poorer linearity of the 5 pJ/K can be the cause of the worse energy resolution.

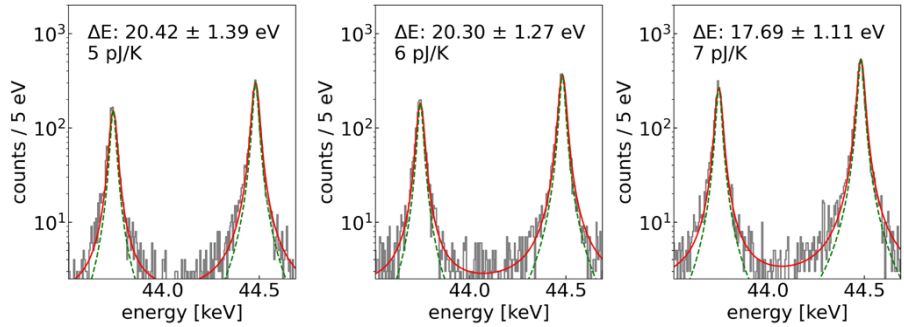

**Fig. 3** Tb K$\alpha_1$, K$\alpha_2$ natural line shape (*dashed green lines*), emission lines (*grey step histogram*) and fittings (*red lines*) of different sizes of pixels.

In order to evaluate the devices' performance at ~ 20 keV, we measure the 17.5 keV Mo K$\alpha_1$ fluorescence spectrum at NIST, and calculate that the $\Delta E$ of the 5 pJ/K and 7 pJ/K pixels are 9.30 ± 0.34 eV and 11.50 ± 0.32 eV, roughly proportional to the square root of $C$. The gain scale in this calculation is determined by a cubic-spline interpolation in the log($E$) vs. log($P$) space. The 6 pJ/K pixel $\Delta E$ was not measured due to a broken channel connection. Given the devices' better energy resolutions at 17.5 keV, greater linearity at lower energies as shown in Fig. 1, and the availability of nearby anchor points, the calibration uncertainties of these devices are expected to be smaller at 20 keV than those at 44 keV. Therefore, the 6 pJ/K and 7 pJ/K pixels satisfy the 20 keV measurement requirement as well.

**4 Conclusion**

We developed TES microcalorimeters for µ-Ar x-ray emission energy measurements at 44 keV and 20 keV. The TESs connect to sidecar absorbers, which are made of Au to set the device heat capacity and thick electroplated Bi bilayers to increase quantum efficiency. The absorbers have heat capacities of 5 pJ/K, 6 pJ/K, and 7 pJ/K. The energy resolutions and calibration accuracies of the different devices have been characterized with fluorescence spectra on individual pixels. We achieve absolute energy

# Title

accuracies of ~ 1 eV on the 6 pJ/K and 7 pJ/K pixels at 44 keV, meeting the requirement of the μ-Ar project. In the next step, we will further increase the thickness of the absorber Bi layer in order to push the quantum efficiency and make a microcalorimeter array with ~ 100 pixels to be used at the Japan Proton Accelerator Research Complex (J-PARC) for the final muonic atom experiment.

**Acknowledgements** This work was partially supported by the Grants-in-Aid for Scientific Research (KAKENHI) from MEXT and JSPS (Nos. 18H05458, 18H03713, 18H03714, 20K20527), and used resources of the Advanced Photon Source, a U.S. Department of Energy (DOE) Office of Science User Facility, operated for the DOE Office of Science by Argonne National Laboratory under Contract No. DE-AC02-06CH11357.